\documentclass[twocolumn]{aastex63}

\usepackage{amsmath}
\usepackage{gensymb}
\usepackage{caption,subcaption}
\usepackage{changepage} 
\usepackage{amssymb}
\usepackage{graphicx}
\usepackage{xcolor}
\usepackage{hyperref}

\usepackage{multirow}
\usepackage{courier} 

\usepackage{enumitem}

\newcommand{\mdot}{\hbox{$\dot{M}$}}

\newcommand{\mbh}{\hbox{$M_{\rm BH}$}}
\newcommand{\msun}{\hbox{M$_{\odot}$}}

\newcommand{\rms}{\sigma^2_{\mathrm{rms}}}

\newcommand{\fscr}{\hbox{$f_{\mathrm{scr}}$}}
\newcommand{\fcol}{\hbox{$f_\mathrm{col}$}}

\newcommand{\kerrbb}{\textsc{kerrbb}}
\newcommand{\simpl}{\textsc{simpl}}
\newcommand{\ztbabs}{\textsc{ztbabs}}

\begin{document}

\title{Disk-to-Corona State Transition and Extreme X-ray Variability in the Tidal Disruption Event  AT2019teq}

\shorttitle{AT2019teq}

\shortauthors{Berger et al.}

\email{vlberger@mit.edu}

\author[0000-0002-7303-8144]{Vera Berger}
\altaffiliation{NSF Graduate Research Fellow}
\affiliation{Department of Physics \& Kavli Institute for Astrophysics and Space Research, Massachusetts Institute of Technology, Cambridge, MA 02139, USA}

\author[0000-0003-0172-0854]{Erin Kara}
\affiliation{Department of Physics \& Kavli Institute for Astrophysics and Space Research, Massachusetts Institute of Technology, Cambridge, MA 02139, USA}

\author[0000-0002-0568-6000]{Joheen Chakraborty}
\affiliation{Department of Physics \& Kavli Institute for Astrophysics and Space Research, Massachusetts Institute of Technology, Cambridge, MA 02139, USA}

\author[0000-0003-4127-0739]{Megan Masterson}
\affiliation{Department of Physics \& Kavli Institute for Astrophysics and Space Research, Massachusetts Institute of Technology, Cambridge, MA 02139, USA}

\author[0000-0002-7226-836X]{Kevin Burdge}
\affiliation{Department of Physics \& Kavli Institute for Astrophysics and Space Research, Massachusetts Institute of Technology, Cambridge, MA 02139, USA}

\begin{abstract}
We present a five-year X-ray spectral and timing analysis of the optically selected Tidal Disruption Event (TDE) AT2019teq, which displays extreme variability, including order-of-magnitude changes in flux on minute-to-day timescales, and a rare late-time emergence of hard X-ray emission leading to the longest-lived corona in a known TDE.
In one epoch, we detect sub-mHz quasi-periodic oscillations with significance tested via MCMC-based red-noise simulations  (p $\leq 0.03$).
AT2019teq exhibits a clear spectral evolution from a soft (blackbody-dominated) state to a hard (power-law-dominated) state, with a late-time radio brightening that may be associated with the state transition. We identify similarities between AT2019teq’s evolution and X-ray binary soft-to-hard state transitions, albeit at higher luminosity and much faster timescales.
We use the presence of both a disk-dominated and a corona-dominated state to apply multiple mass estimators from X-ray spectral and variability properties.
These techniques are mutually consistent within $2\sigma$ and systematically yield a lower black hole mass ($\log(\mbh/\msun)=5.67\pm0.09$) than inferred from host galaxy scaling ($\log(\mbh/\msun)=6.14\pm0.19$).
\end{abstract}


\section{Introduction}\label{sec:intro}
Tidal disruption events (TDEs) occur when a star passes too close to a massive black hole (MBH) and is torn apart by tidal forces. 
A fraction of the stellar debris remains bound and falls back, which causes a bright accretion flare \citep{rees88, evans89, gezari21}.
TDEs provide a window into otherwise quiescent black holes and allow the study of MBH accretion disk formation and evolution over observable timescales.

The first TDE candidates were discovered in the soft X-rays with ROSAT \citep{bade96}.
X-ray spectra of these events are typically extremely soft and quasi-thermal, with blackbody temperatures of $kT \sim 30$–60~eV \citep{ulmer99, auchettl17},  thought to originate from a newly formed accretion disk.
The soft X-ray light curves of TDEs show a diversity of phenomena.
Some track a smooth decline similar to the optical \citep[e.g., ASASSN-14li,][]{holoien1614li}.
Others show much stronger and more rapid X-ray variability that does not decline as precipitously as the optical/UV emission \citep{vanvelzen21_17tdes, wevers21, yao24}.

TDEs do not typically exhibit hard X-ray coronae \citep{komossa15}  that are ubiquitous in Active Galactic Nuclei \citep[AGN,][]{kara25}. 
In AGN, the corona is characterized by a compact, optically thin plasma with temperature $\sim 10^9$~K that produces hard X-rays.
Some TDEs with soft X-ray spectra have exhibited a weak hard tail \citep{holoien16, saxton17, kara18}, and a few have  transitioned to a hard power-law-dominated state \citep{wevers21, yao22, guolo24}. 
This hardening, first observed within 100-200~days of optical peak, has been interpreted as the emergence of a magnetically-dominated corona \citep{yao22, guolo24}.

Observed state transitions in TDEs  may provide a unique probe of the formation and evolution of disks, coronae and jets in MBHs.
The scaling of accretion state transitions with black hole mass remains a fundamental question.
State transitions are ubiquitous in stellar-mass black hole X-ray binaries (XRBs). XRB outbursts follow a canonical hysteresis pattern in the X-ray hardness--intensity diagram, where sources brighten as they transition from hard to soft states, then dim and harden as they return to quiescence \citep{remillard06, wang22}.
Many BHXRBs show strong radio--X-ray coupling, with compact radio jets present in the hard state that are then quenched in the soft state. 
This coupling appears to scale with mass, defining a ``Fundamental Plane" of black hole activity that links radio luminosity, 2--10~keV X-ray luminosity, and black hole mass \citep[e.g.,][]{merloni03, falcke04}.
While AGN luminosities align with the Fundamental Plane, comparable changes in accretion rate to XRB outbursts are expected to occur over $\gtrsim 10^3$--$10^6$~years, far beyond observable timescales. 
TDEs thus provide a rare opportunity to test how accretion state transitions scale with black hole mass.

Accurate estimates of the black hole mass in TDEs are essential to constrain the scaling of accretion physics, determine spin and correlate with host galaxy properties.
TDEs naturally probe the low end ($\sim10^5$--$10^8~\msun$) of the SMBH mass spectrum, since there is thought to be an upper limit to the mass at which a black hole can tidally disrupt a star outside the event horizon \citep{hills75, rees88, vanvelzen18}. 
Host galaxy scaling relations are not well constrained at low mass \citep[e.g.,][]{haring04, gultekin09, kormendy13}, and direct methods such as reverberation mapping for AGN are not applicable to TDEs due to the absence of a virialized broad line region.

Efforts to estimate $\mbh$\, with UV/optical light curve properties have shown varying consistency with host galaxy scaling relations \citep{mockler19, ryu20, hammerstein23}. 
The models rely on uncertain assumptions about origins of early-time emission. MOSFiT, for instance, assumes reprocessing and rapid circularization \citep{mockler19}, while TDEMass assumes shocks between debris streams and slow circularization \citep{ryu20}.
Late-time UV/optical emission is thought to be disk-driven, and recent work has found promising correlations between late-time light curve properties and black hole mass \citep{mummery24}. 
In this work, we employ X-ray based techniques to estimate the black hole mass and compare our results to host galaxy scaling relations and UV/optical-based methods.

AT2019teq (18:59:05.498, +47:31:05.66) was discovered on October 20, 2019 by ZTF at a redshift of 0.0878.
Optical spectroscopy classified the event as a TDE-H+He \citep{hammerstein23}.
Follow-up with \textit{XMM-Newton} and Swift revealed a soft, thermal X-ray spectrum. 
Three years later, an \textit{XMM-Newton} observation 
revealed that AT2019teq had brightened and hardened in the X-rays \citep{yao22atel}.
Contemporaneous VLA observations detected radio emission at 6~GHz which subsequently faded, consistent with an outflow that peaked between 400-1000~days post optical maximum \citep{cendes22, cendes24}.

Here we present a five-year X-ray spectral and timing analysis of AT2019teq. 
The source displays rapid high-amplitude variability, with order-of-magnitude changes in flux on timescales from minutes to days.
AT2019teq exhibits a state transition from an optically bright, X-ray dimmer soft state to a hard and more X-ray-luminous state within three years of optical peak. 
The source has remained uniquely bright and hard in the X-rays, providing a rare opportunity to trace the long-term evolution of both an accretion disk and a corona in a massive black hole.
We use the presence of both disk-dominated and corona-dominated states to estimate the black hole mass with techniques from both regimes.

In Section \ref{sec:data}, we describe the observations and data reduction.
In Section \ref{sec:results}, we present AT2019teq's spectral evolution and timing properties which lead to independent mass estimates. 
Section \ref{sec:discussion} compares mass constraints on the system from X-ray and UV/optical methods, and discusses AT2019teq's accretion states in the context of XRBs.
Throughout this work we assume a standard $\Lambda$CDM cosmology with $H_0=70$~km\,s$^{-1}$\,Mpc$^{-1}$, $\Omega_m=0.27$ and $\Omega_\Lambda=0.73$.

\section{Observations and data reduction}\label{sec:data}

\begin{figure*}[ht]
    \centering
    \includegraphics[width=0.8\linewidth]{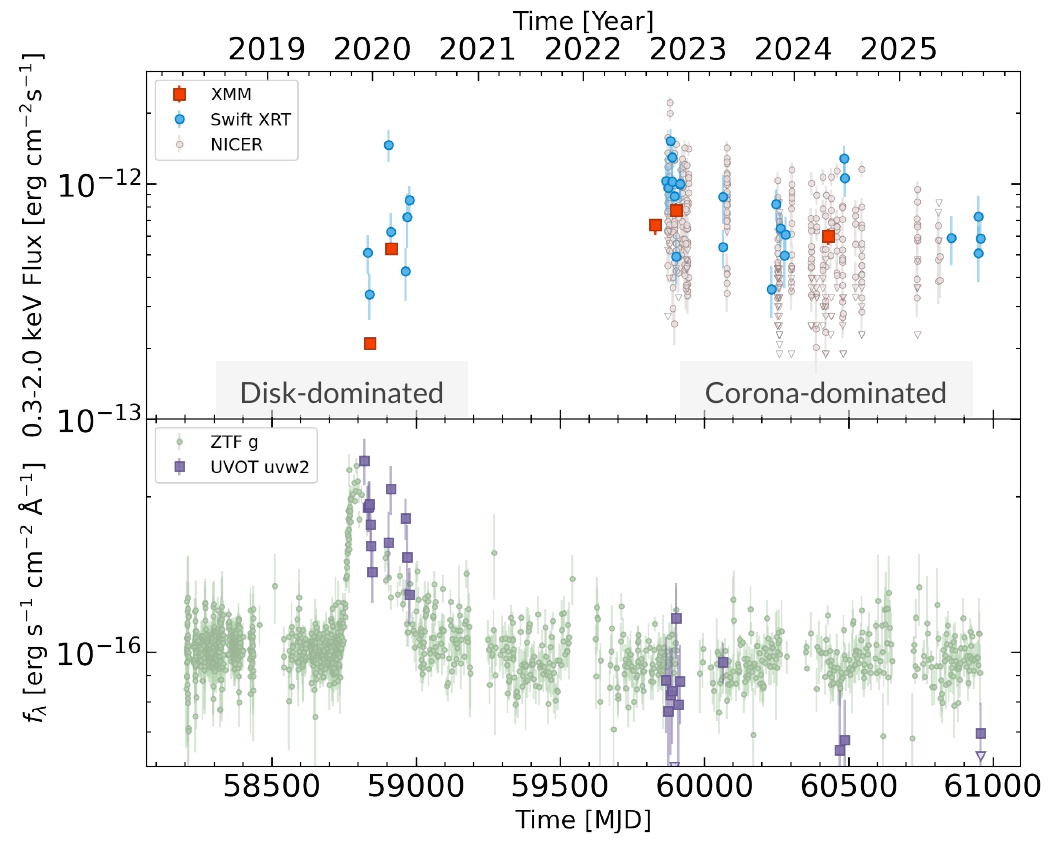}
    \caption{Long term X-ray (upper)  and UV/optical (lower) light curves of AT2019teq. X-ray flux is computed between  $0.3$--$2$~keV for \textit{XMM-Newton} (red), Swift XRT (blue), and NICER  (beige). The ZTF $g$-band light curve is computed from difference photometry, 
    converted to total flux using a reference magnitude and
    expressed at the pivot wavelength.  The Swift UVOT uvw2 light curve is shown in purple.}
    \label{fig:lcs_all}
\end{figure*}

\subsection{\textit{XMM-Newton}}
\textit{XMM-Newton} observed AT2019teq five times between December 2019 and April 2024 for a total of 169~ks.
The first two observations took place during the brightest optical flare, and the remaining three observations took place after the optical had significantly faded.
Observation details and exposure times are listed in Table \ref{tab:obs_log}. 
We focus on observations from the pn instrument on the European Photon Imaging Camera \citep[EPIC-pn,][]{struder01} for its high sensitivity relative to the MOS cameras.
We use standard data reduction procedures with the \textit{XMM-Newton} Science Analysis System (SAS) v20.0.0 and Heasoft v6.32.
Events are restricted to singles and doubles (\texttt{PATTERN <= 4}).
The source is extracted from a region with a $25\arcsec$ radius, and the background  is extracted from a region with a $50\arcsec$ radius on the same detector.

All five observations except XMM4 were impacted by strong background particle flaring.
These observations are therefore processed with a more permissive background threshold of $10$~ct~s$^{-1}$, chosen by visual inspection to remove the most extreme flares while maximizing usable exposure time.
To ensure this high threshold does not introduce contamination, we verify that no residual flaring signatures are visible in the background-subtracted light curves.
XMM4, which was unaffected by background flaring, exhibits high variability consistent with the remaining observations, which supports that the observed variability is intrinsic to the source.
Durations of the cleaned light curves are recorded in Table \ref{tab:obs_log}, and 10--12~keV particle background light curves are in Appendix \ref{app:bg}.

\begin{figure*}
    \centering
    \includegraphics[width=0.85\linewidth]{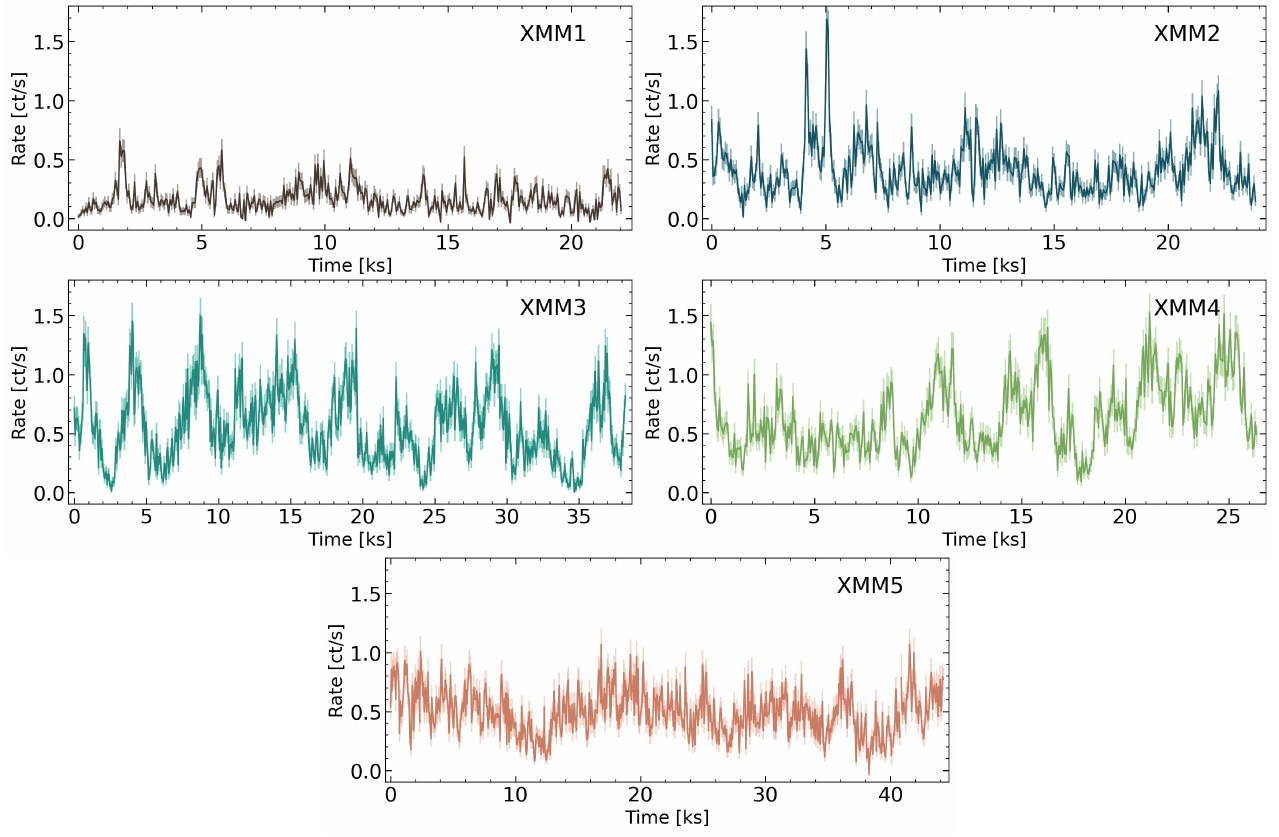}
    \caption{\textit{XMM-Newton} 60~s light curves for each observation. Light curves are extracted from 0.3--7.0~keV, except for the first two observations which are truncated at 2.0 and 4.0~keV, respectively, due to background domination at higher energies.
    The light curves show extreme variability, with order-of-magnitude changes in flux on timescales as short as 10 minutes.}
    \label{fig:lcs_xmm}
\end{figure*}

The 0.3--2~keV \textit{XMM-Newton} light curve binned by observation is shown in Figure \ref{fig:lcs_all}.
AT2019teq brightens from the first \textit{XMM-Newton} observation, then the soft X-ray flux stays relatively constant over time.
However, the source is highly variable within individual exposures.
Figure \ref{fig:lcs_xmm} shows 60~s light curves for each XMM observation.
There are order-of-magnitude changes in flux on timescales as short at 10 minutes. 
The light curve for XMM3 appears to show some periodic behavior every $\sim3$~ks; we characterize this quasi-periodicity in Section \ref{sec:qpo}.

Finally, spectra are grouped to a minimum of 25 counts per bin.
We fit each spectrum in the energy band where the signal dominates over the noise. This corresponds to 0.3--2~keV for XMM1, 0.3--4.0~keV for XMM2, 0.3--7~keV for XMM3 and XMM5, and 0.3--8~keV for XMM4.
The spectra are shown in Figure \ref{fig:spectra}, binned for visual clarity. 
The source exhibits visible X-ray brightening and hardening over time, which we discuss in Section \ref{sec:spectra}.

\subsection{NICER}
The Neutron star Interior Composition Explorer \citep[NICER,][]{gendreau16} observed AT2019teq between October 2022 and June 2025 for a total of 107~ks. 
We follow the method outlined in \citet{chakraborty24} to obtain light curves for faint sources with variable background; we discuss it briefly here.
We first process observations with \texttt{nicerl2} and generate GTI files between 150-200~s in length with \texttt{nimaketime}.
Spectra are extracted for individual GTIs with \texttt{nicerl3-spec}, and fit with \textsc{tbabs$\times$simpl$\times$diskbb} and the SCORPEON background model.
The 0.3-2.0~keV flux is computed for each GTI and plotted in Figure \ref{fig:lcs_all}. 
The source shows order-of-magnitude variability across GTIs on sub-day timescales.

\subsection{Swift}
\subsubsection{XRT}
We obtain Swift-XRT data products using an online tool from the UK Swift Science Data Centre \citep{evans07, evans09}. 
Light curves are computed in the 0.3--2~keV band for direct comparison to NICER and \textit{XMM-Newton}.
Count rates are converted to flux assuming a constant conversion factor from spectral fitting;
we combine eight observations (2022-10-15 to 2022-12-03, 13.8 ks total) to optimize signal-to-noise and fit a power-law model with a Galactic absorber fixed at $4.54\times10^{20}$~atoms/cm$^2$ \citep{h14pi}.
We determine a counts-to-flux conversion factor of $2.14\times10^{-11}$ erg cm$^{-2}$ ct$^{-1}$.
Figure \ref{fig:lcs_all} displays the Swift XRT light curve, binned by snapshot, in blue.

\subsubsection{UVOT}
We use observations from the UVOT uvw2  filter ($\lambda_\mathrm{ref}=2023.44$~\AA), the bluest among UVOT filters and therefore expected to have the lowest level of host galaxy contamination.
We use standard processing for UVOT data products with Heasoft.
The source is extracted from a circular region with 5\arcsec\, radius, and the background is extracted from a nearby region with 15\arcsec\, radius.
Source and background regions are fixed across all observations.
We use \texttt{uvotimsum} to co-add the individual image extensions within each observation, and perform photometry with \texttt{uvotsource}. 
The Swift UVOT uvw2 light curve is shown in the lower panel of Figure \ref{fig:lcs_all} (purple squares).

\subsection{ZTF}
We obtain a ZTF light curve using forced photometry, which uses difference imaging to measure fluxes at fixed positions rather than requiring $\sim5\sigma$ detections per standard ZTF processing. 
The forced photometry approach allows for detection of fainter sources and avoids confusion in crowded fields.
Difference fluxes are converted to absolute fluxes by adding the source flux from a deep reference image. 
We then correct for zeropoint offsets and convert to physical units using the $g$-band pivot wavelength \citep[$\lambda_{p}=4783.50$~\AA, ][]{rodrigo12}.
The ZTF $g$-band light curve is shown in the lower panel of Figure \ref{fig:lcs_all}.
\citet{langis25} claim a precursor flare detection for AT2019teq around MJD 58475, presumably using standard ZTF photometry. 
This feature is not recovered in forced photometry or the ATLAS light curve, which suggests the precursor flare is an artifact.

\begin{table*}[ht]
\centering
\caption{\textit{XMM-Newton} observation log}
\begin{tabular}{llllll}
\hline\hline
Epoch & OBSID    & Start Date &  Exposure  & Cleaned & Total Counts \\
 &     &   &  [ks]  & Exp. [ks] &  0.3--2.0~keV [$10^3$ cts] \\
\hline
XMM1  & 842591701 & 2019-12-24 & 25.6        &  22.0   &     3.6      \\
XMM2  & 842592401 & 2020-03-07 & 33.6        &  23.9      &    9.5    \\
XMM3  & 902760901 & 2022-09-08 & 54.1       & 38.3       &   19.0  \\
XMM4  & 913992001 & 2022-11-19 & 30.0       & 26.4      &     14.9   \\
XMM5  & 935190101 & 2024-04-30 & 55.7       &   44.3      &     19.6    \\ \hline
\end{tabular}
\label{tab:obs_log}
\end{table*}

\section{Results}\label{sec:results}
In this section we present spectral, root mean square (rms) variability and power-spectral analyses based on the five \textit{XMM-Newton} observations. Each leads to an estimate of the black hole mass, which we will discuss in Section \ref{sec:discussion}.

\subsection{Spectral evolution}\label{sec:spectra}
We perform spectral modeling of the five XMM observations with \textsc{xspec}  v12.13.1 \citep{arnaud96} using $\chi^2$ statistics.
Models are convolved with  \textsc{tbabs} to account for Galactic hydrogen absorption and \textsc{ztbabs} for absorption at the host redshift \citep{wilms00}. 
The Galactic hydrogen column density is fixed at $4.54\times10^{20}$~atoms/cm$^2$ \citep{h14pi}, and the column density at the host redshift is left as a free parameter.

Spectra are fit with \simpl$\times$\kerrbb\, to describe the thermal disk and Comptonized corona emission.
The thermal emission is modeled using \kerrbb, which considers a geometrically thin, optically thick relativistic accretion disk around a Kerr black hole \citep{li05}.
The disk emits locally as a blackbody with a constant color correction factor \fcol\, to account for electron scattering and Comptonization.
We adopt the default zero-torque inner boundary condition $\eta=0$.

Since a disk continuum has only two independent observables (temperature scale and flux amplitude), disk models have intrinsic degeneracies and typically constrain only two parameters
\citep[e.g.,][]{li05, salvesen21}.
In this study, we are most interested in mass and accretion rate,
and so freeze the color correction factor ($\fcol=2.4$), disk inclination ($i=45\degree$) and spin ($a=0.999$). 
We choose $\fcol=2.4$ based on theoretical  disc models for a $10^6~\msun$ black hole with a high accretion rate \citep{ross92, done12}.
There is uncertainty surrounding the color correction factor, with a possible range of $\approx1.4-2.5$ \citep[e.g.,][]{li05}. 
We explore this range and find that lowering \fcol\, only decreases the mass estimate, by at most a factor of three for the lowest tested value of $\fcol=1.4$.
The disk inclination is positively degenerate with mass, and the spin is negatively degenerate with mass.
We further discuss and the influence of these parameters on inferred black hole mass in Section \ref{sec:disc_kerrbb}. 

\simpl\,  generates a power-law via Compton scattering of a fraction of disk photons \citep{steiner09}. 
We fit for the power-law index $\Gamma$ and scattered fraction $f_{\mathrm{scr}}$, restricting the model to up-scattering only. 

To constrain a black hole mass, we fit all five observations simultaneously. 
We attempted a fit where the extragalactic hydrogen column density, photon index $\Gamma$ and scattering fraction \fscr\, were allowed to vary between observations, but the SNR of our data was insufficient.
Because there is no evidence of changes in obscuration, we tie the host hydrogen column density across observations. 
Moreover, because the power-law component is very weak at early times, we cannot constrain both the photon index and \fscr, and so we tie the photon index across all observations, leaving only \fscr\, to vary.
We also tie the mass, which should remain constant across observations, and allow
the accretion rate \mdot\ to vary.

Final fit parameters and uncertainties are determined from MCMC analysis with \texttt{xspec\_emcee} \citep{sanders18}.
We use 5000 iterations (500 burn-in) and 50 walkers. 
Table \ref{tab:spectral_fits} displays the median fit parameters and $1\sigma$-uncertainties. 
The inferred extragalactic hydrogen column density is $0.08\pm0.01 \times 10^{22}$ atoms cm$^{-2}$.
The inferred photon index (driven largely by XMM3,4,5) is $2.84\pm 0.02$, and the inferred black hole mass is  $3.0^{+0.4}_{-0.3}\times10^5$~\msun.

 \begin{table*}[ht]
 \centering 
   \caption{Best-fit spectral parameters and $1\sigma$ uncertainties for a simultaneous fit to five XMM observations with \textsc{tbabs$\times$ztbabs$\times$simpl$\times$kerrbb}. The host galaxy hydrogen column density, photon index, and mass are tied across observations.}
     \begin{tabular}{llllllll} 
     \hline \hline 
    Parameter [Units] & XMM1 & XMM2 & XMM3 & XMM4 &XMM5 \\ \hline 
    \textsc{ztbabs nH} [$10^{22}$ atoms cm$^{-2}$] & $0.08 \pm 0.01$ & — & —& —& — \\ 
    \simpl\, $\Gamma$ & $2.84 \pm 0.02$ & — & —& —& — \\ 
     \kerrbb \, M [M$_\odot$] & $3.0^{+0.4}_{-0.3}\times10^5$ & — & —& —& — \\ 
    \kerrbb \, $f_{scr}$ & $0.04\pm0.02$ & $0.14\pm0.01$ & $0.83\pm0.07$ & $0.86\pm0.06$ & $0.96^{+0.03}_{-0.05}$ \\ 
    \kerrbb \, $\dot M$ $[M_\odot$yr$^{-1}]$ & $0.0012\pm0.0001$ & $0.0021\pm0.0002$ & $0.0013\pm0.0001$ & $0.0015\pm0.0001$ & $0.0011\pm0.0001$ \\ 
    $\chi^2$/dof & 45.2/41=1.10 & 91.8/81=1.13 & 166.4/129=1.29 & 152.7/138=1.11 & 113.1/101=1.12 \\ 
    0.3--10~keV Luminosity [$10^{43}$ erg s$^{-1}$] & $0.44\pm0.02$ & $1.23\pm0.02$ & $1.97\pm0.18$ & $2.29\pm0.22$ & $1.75\pm0.18$ \\
    \hline 
 \end{tabular} 
  \label{tab:spectral_fits} 
 \end{table*}

 \begin{figure*}
    \centering
    \includegraphics[width=0.6\linewidth]{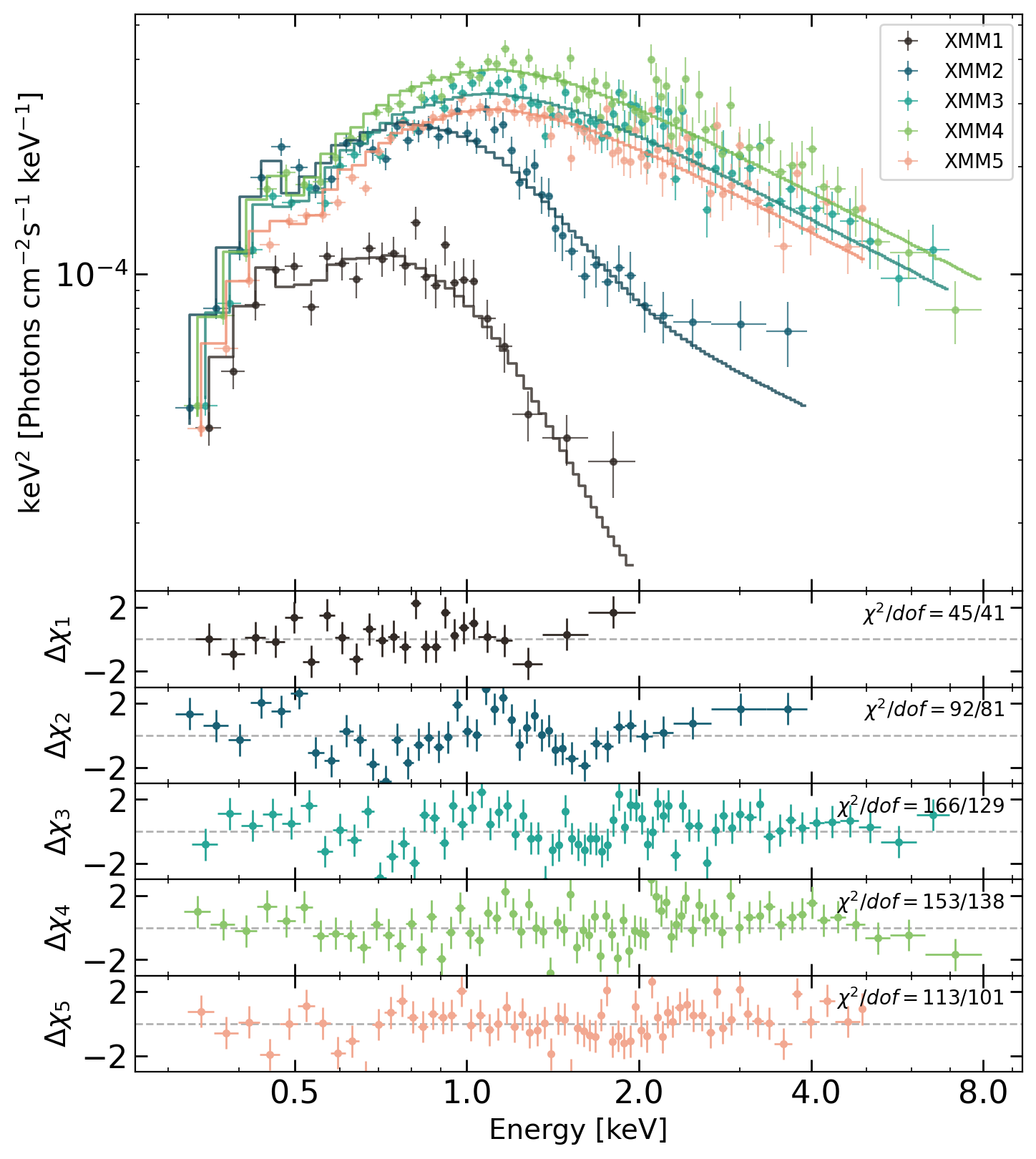}
    \caption{\textit{Upper panel:} \textit{XMM-Newton} EPIC-pn unfolded spectra and best-fit models for all observations. Color lightens with time. \textit{Lower panels:} Normalized residuals $\Delta\chi =$(data - model)/error by observation. Spectra are binned for visual clarity.}
    \label{fig:spectra}
\end{figure*}

Figure \ref{fig:spectra} displays the resulting fits and normalized residuals.
The X-ray spectra show a clear spectral hardening and brightening over time.
The \simpl$\times$\kerrbb\, model provides a good fit ($\chi^2/$d.o.f.$<1.15$) to all epochs except XMM3, with $\chi^2/$d.o.f.$=166.4/129=1.29$.
XMM3 and XMM2 show a visible residual excess around 0.7--1~keV.
Both observations' fit statistics are improved by the addition of a Gaussian emission or absorption component.
XMM3's spectrum is equally well fit by a gaussian emission line at 1~keV and an absorption line at 0.75~keV, with a $\Delta \chi^2=11$ for three degrees of freedom.
XMM2's spectrum is also improved by a gaussian emission line at 1~kev with  $\Delta \chi^2=12$,  or an absorption line at 0.75~keV with  $\Delta \chi^2=17$.
Such features, both in emission and absorption, have been interpreted as due to mildly relativistic accretion disk outflows 
\citep[e.g.,][]{kara18, masterson22, pasham24}.
As the detections here are marginal, we do not attempt more physically motivated photoionization models at this time.
While XMM3 is the only observation appearing to require an additional spectral component, it has the highest total counts (along with XMM5), suggesting that a feature may be less pronounced in other observations due to lower SNR.

\begin{figure}
    \centering
    \includegraphics[width=\linewidth]{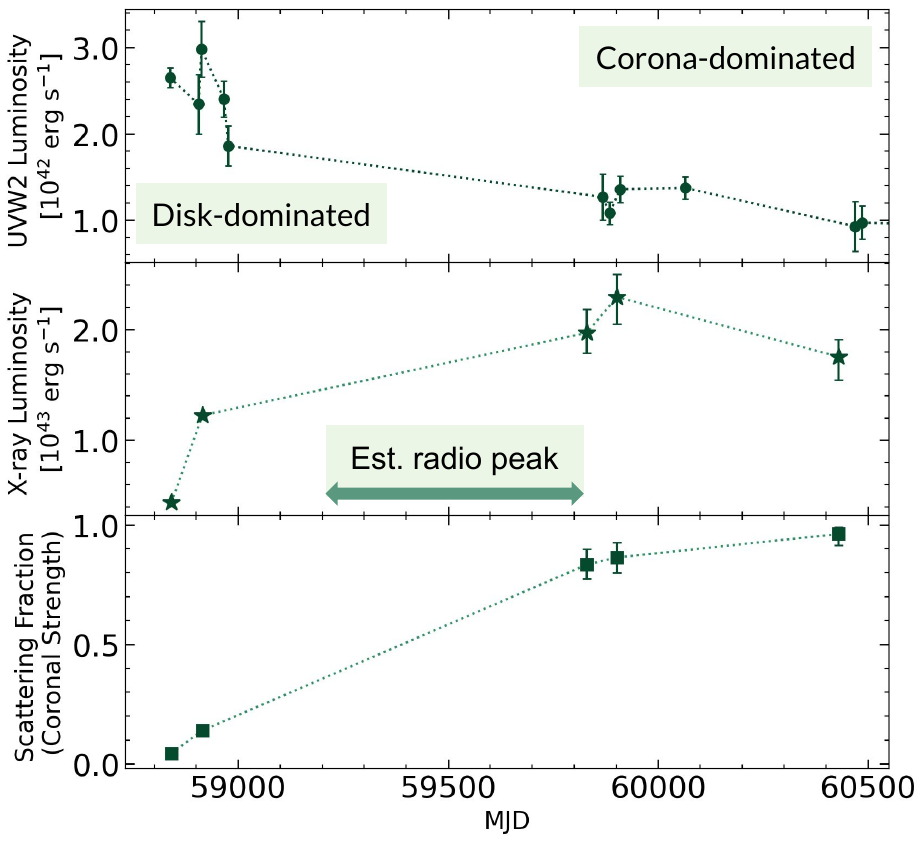}
    \caption{Time evolution of luminosity and coronal strength. \textit{Top to bottom:} UVOT uvw2 luminosity (circles), X-ray luminosity (0.3--10~keV, stars) and scattering fraction (squares). Declining radio emission was detected around MJD 59900 after non-detections in early times, consistent with an outflow that peaked $\sim400$--$1000$~days post optical peak \citep{cendes22}.}
    \label{fig:spectral_params}
\end{figure}

The system evolves from an X-ray-fainter and soft blackbody-dominated state in early times to a brighter and harder power-law-dominated state in later observations.
Figure \ref{fig:spectral_params} shows the time evolution of the UVOT uvw2 luminosity, 0.3--10~keV luminosity and scattering fraction.
The UV luminosity decreases over time as the initial flare subsides.
The X-ray luminosity increases by a factor of $2.8$ between the first two epochs, and stays relatively constant through the remaining XMM observations. 
The fraction of Compton-upscattered photons \fscr\, rises monotonically from $0.04\pm0.02$ in December 2019 to $0.96^{+0.03}_{-0.05}$ in April 2024, indicating a transition to a Comptonized, hard X-ray state.
A hard excess is already present by XMM2, only 121 days after the optical peak.
However, the substantial gap between the early- and late-time observations of more than two years limits constraints on when and how rapidly the state transition happened.

AT2019teq does not return to a soft state in our final \textit{XMM-Newton} observation.
In fact, Swift XRT spectra from October 2025 confirms the hard state has persisted for at least 1100 days, indicating the longest-lived corona of any known hard-state TDE.

\subsection{Timing analysis}\label{sec:qpo}

\subsubsection{Excess variance}

The normalized excess variance $\rms$ provides a  measure of the intrinsic variability amplitude of a light curve \citep[e.g.,][]{nandra97, edelson02}. 
For a light curve with variance $S$ and mean square error $\sigma_\mathrm{err}$, the normalized excess variance is given by
\begin{equation}
\rms = \frac{S^2-\overline{\sigma_\mathrm{err}^2}}{\overline{x}^2},
\end{equation}
where  $\overline{x}$ is the mean count rate and $\overline{\sigma_\mathrm{err}^2}=\frac{1}{N}\sum_{i=1}^N\sigma_i^2$ is the mean square error.
$\rms$ is equivalent to the integral of the power spectral density for large $N$, and therefore measures the fractional variability over the frequency range covered by the light curve \citep{vaughan03}.
AGN studies routinely use $\rms$ since it can be measured from short observations without continuous sampling, making it a practical alternative to full PSD analysis \citep{ponti12}.

\citet{ponti12} constructed empirical relations between $\rms$ and black hole mass for a sample of 32 AGN with masses determined from reverberation mapping. 
These relations allow $\mbh$ to be estimated from short-term X-ray variability, the general idea being that the relevant physical timescales (e.g. light crossing, dynamical, viscous timescales) all scale linearly with mass, and thus, for a given variability timescale, higher-mass black holes will exhibit less variability.
Excess variance is computed from 2--10~keV light curves, and relations are reported for a range of durations: 10, 20, 40, and 80~ks. 
We use the 20~ks relation to simultaneously maximize the number of usable observations and segment exposure time.

The $\rms$ calculation is restricted to the later \textit{XMM-Newton} observations, since these probe AT2019teq's corona-dominated state and since XMM1–2 become background dominated between 2--4~keV.
The 20~ks light curves are sampled at 200~s cadence and contain no gaps. XMM5 is sufficiently long to extract two 20~ks segments, and the others yield one each.
We compute an unweighted mean $\rms = 0.22 \pm 0.01$ across XMM3,4,5, which, according to the \citet{ponti12} AGN relation, would suggest a black hole mass of $6.4 \pm 2.4 \times 10^5\,\msun$.

The $\rms$ decreases from $0.31\pm0.03$--$0.36\pm0.03$ in XMM3,4 to $0.09\pm0.02$--$0.13\pm0.03$ in XMM5. 
Given the limited number of observations, it is unclear whether this reflects a long-term decrease in variability.
The nearly identical spectra across the three observations suggest they sample the same state.
That said, the lowest measured excess variance $\rms=0.09\pm0.02$ in XMM5 provides a strong upper limit on the black hole mass of $1.4\pm 0.5 \times 10^6\,\msun$. 
Further discussion of the applicability of this method for TDEs is given in Section \ref{sec:disc_rms}.

\subsubsection{PSD analysis}
Motivated by the seemingly periodic modulation seen in XMM3, we perform a power spectral density analysis on all observations to statistically test for the presence of a quasi-periodic oscillation (QPO). 
We perform the PSD analysis on the 0.3--4~keV light curve in order to maximize signal-to-noise, and bin the light curve to 30~s. 
We follow the procedure outlined in  \citet{masterson25} to fit the PSDs and estimate the significance of a potential QPO feature.
In order to ensure that our QPO statistic is not biased by our choice of broadband noise model, we start by fitting unbinned PSDs with two separate broadband noise models (with an added constant for Poisson noise). 
The power-law $\mathcal P(f) = Nf^{-\alpha}+c$ is widely used for XRB and AGN power spectra.
We also use a Lorentzian centered at $\nu_0=0$:
\begin{equation}
    \mathcal P(f) = \frac{2r^2 \Delta}{\pi} \frac{1}{\Delta^2+(\nu - \nu_0)^2}+c.
\end{equation}

\begin{figure}
    \centering
    \includegraphics[width=0.95\linewidth]{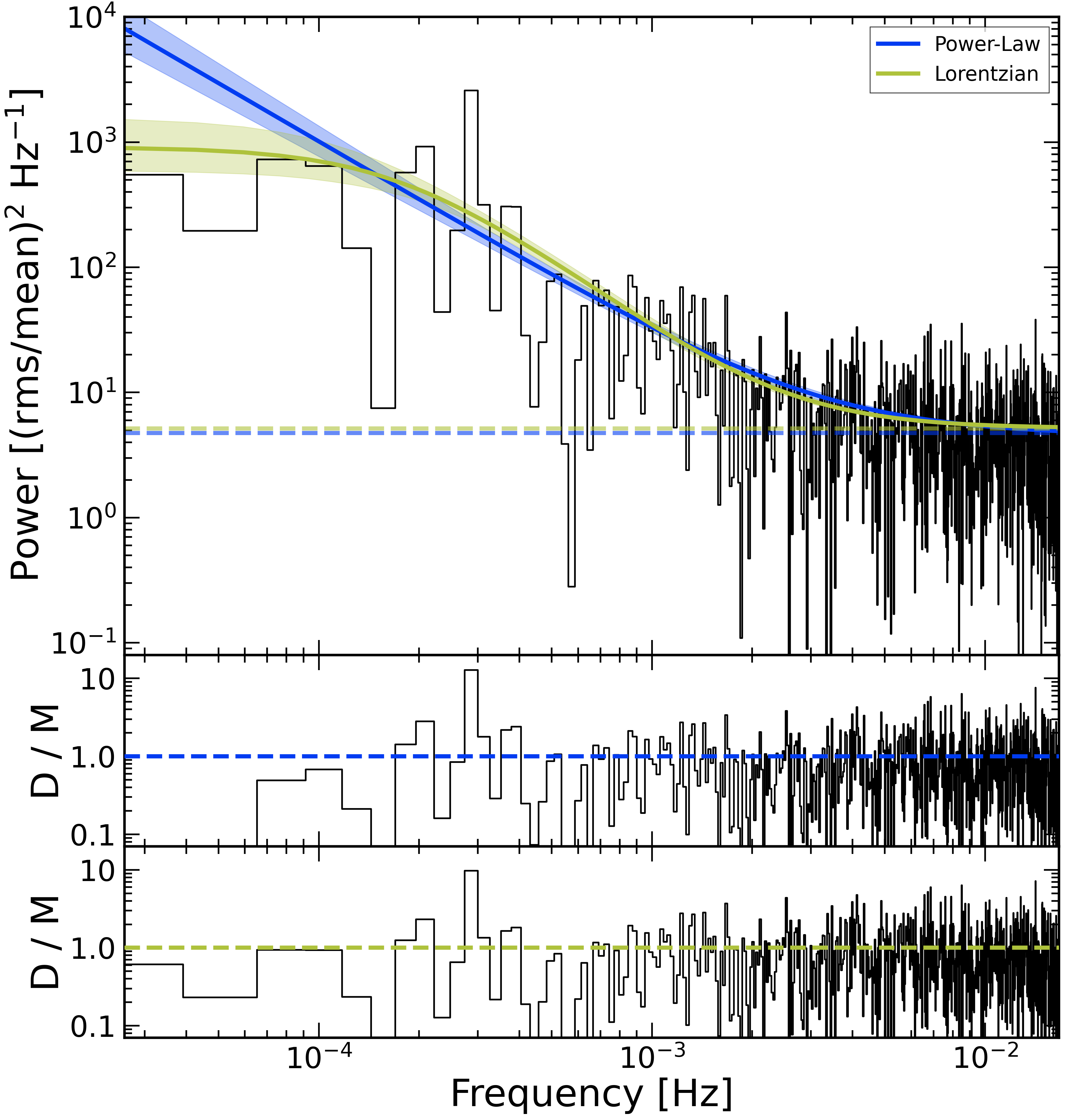}
    \caption{
    \textit{Upper panel:} Unbinned power spectrum with best-fit broadband noise models.  Solid lines show power-law (blue) and Lorentzian (green) models, with shading for $1\sigma$ uncertainties. 
    Horizontal lines denote the Poisson noise level. 
    \textit{Lower panel:} Data/model ratio with horizontal dashed lines at 1.}
    \label{fig:qpo_psd}
\end{figure}

PSD continuum fits are computed using Markov chain Monte Carlo (MCMC) analysis with \texttt{emcee} \citep{foreman-mackey13}.
The MCMC is initialized with fit parameters from maximum likelihood estimation.
We use 32 walkers with 55,000 steps (5000 burn-in) and uniform priors.
All observations except XMM3 are consistent with only red noise. 
Figure \ref{fig:qpo_psd} shows the PSD of XMM3 with a prominent, narrow peak above the continuum around $3\times10^{-4}$~Hz.
The power-law and Lorentzian continuum fits both underpredict this feature, with the data/model ratio reaching $12.85$ and $9.7$, respectively.

To characterize the significance of this feature, we use a test statistic $T_R=\max(R_\nu) = 2I_\nu/\mathcal P_\nu$, where $I_\nu$ is the observed power and $P_\nu$ is the model power \citep{vaughan05}.
This statistic is sensitive to single outliers, which is ideal for narrow features (note this differs from \citet{masterson25}, which characterizes a broad QPO spanning multiple frequency channels). 
We compute $T_R$ of $25.7$ and $19.4$ for the power-law and Lorentzian broadband fits. 

We estimate the probability of detecting a feature of this strength by conducting red noise simulations \citep{timmer95}.
Light curves are generated with equivalent mean, $\rms$ and cadence to our observed data.
PSDs are computed and fit to broadband noise models as above.
At high frequencies, the PSD is dominated by Poisson noise, and most outliers in the simulated PSDs appear in this regime.
We therefore calculate $T_R$ at frequencies where the power is above the Poisson noise level.
We implement a cutoff frequency where power-law and noise components contribute equally to the total power.
For XMM3, this corresponds to 2.6 and 2.4~mHz for the power-law and Lorentzian models,
which correspond to Poisson noise levels of 4.8 and 5.2 (rms/mean)$^2$~Hz$^{-1}$, respectively.
These agree well with the theoretical estimate of Poisson noise based on light curve properties: $P_N = 2(\langle x\rangle + B)/\langle x\rangle^2=4.4$~(rms/mean)$^2$~Hz$^{-1}$, where $\langle x\rangle$ and $B$ are the mean source and background count rates \citep{vaughan05}.

Table \ref{tab:qpo} lists the p-values from red-noise simulations of 50,000 light curves. 
For the full PSD, we find $p=0.001$ for the power-law model and $p=0.03$ for the Lorentzian model. 
Upon restricting to the PSD region not dominated by Poisson noise, the p-values decrease to $p_\mathrm{R}=1.5\times10^{-4}$ and $p_\mathrm{R}=3\times10^{-3}$, respectively. 
These results indicate that the QPO feature in XMM3 is statistically significant in both continuum models, with higher significance when the Poisson noise regime is excluded.

We further test QPO significance by adding a Lorentzian component to each broadband PSD model to represent the QPO.
We assess whether this model improves over the continuum-only fits using the Akaike Information Criterion (AIC), with $\Delta$AIC $\gtrsim 10$ typically used to indicate a statistically significant improvement.
Fits are performed as above with MLE-initialized MCMC, using log-uniform priors on the QPO width $\Delta$ to probe its full dynamic range.
The additional feature substantially improves the power-law broadband model ($\Delta$AIC$=17.24$) and marginally improves the Lorentzian model ($\Delta$AIC$=9.89$).
The increased significance in the power-law+QPO model may reflect its improved fit to low-frequency flattening in the PSD, which may not be intrinsic to the QPO.

The power-law and Lorentzian models yield consistent QPO frequencies of $f_{\rm QPO}=0.23\pm0.07$~mHz and $f_{\rm QPO}=0.29\pm0.01$~mHz, respectively.
However, the power-law fit favors a broad feature with a poorly constrained quality factor $Q=1.6^{+22.4}_{-1.1}$, while the Lorentzian fit favors a narrower QPO with $Q=39^{+61}_{-32}$ (Table~\ref{tab:qpo}). 
Both fits show large uncertainties in $Q$, and MCMC analysis reveals no significant correlations between parameters.
As shown in Figure \ref{fig:qpo_psd}, the PSD spike is visibly confined to a single frequency channel, which is consistent with a narrow, low-significance feature.
The width of a single frequency channel sets a lower limit on $Q$ of $11.0$.

Quasi-periodic modulation is evident in the light curve and persists even under strictly periodic assumptions, as illustrated by the periodic shading in Figure~\ref{fig:qpo_lc}. 
The phase-folded light curve, shown in the lower panel, displays a coherent feature that is well fit by a sinusoid.
While QPOs can be transient, detections in follow-up observations would increase confidence.
Because the statistical significance is modest, model-dependent, and confined to a single observation, we classify this as a marginal QPO detection.

\begin{table*}[ht]
\begin{tabular}{lllllll}
\hline\hline
Continuum  & $f_\mathrm{qpo}$                                    & $p_\mathrm{R}$          & $p_\mathrm{R}$  & $p_\mathrm{AIC}$           & Q                    & RMS              \\
      & {[}mHz{]}              &                        &   [$\nu<\nu_{\mathrm{cutoff}}$]                 &                     &                      & {[}\%{]}         \\ \hline
Power-law  & $0.23\pm 0.07$ & $1\times10^{-3}$ & $1.5\times10^{-4}$ & $1.81\times10^{-4}$  & $1.6^{+22.4}_{-1.1}$ & $48 \pm 11$      \\
Lorentzian & $0.29 \pm 0.01$        &  $3\times10^{-2}$ & $3\times10^{-3}$   & $7.11\times10^{-3}$ & $39^{+61}_{-32}$     & $35\pm 19$ \\ \hline
\end{tabular}
\caption{QPO parameters and significance measures for XMM3. From left to right, columns list the continuum model, QPO centroid frequency $f_\mathrm{qpo}$, p-values from red-noise simulations for the full PSD and for frequencies below the Poisson cutoff $p_\mathrm{R}$, p-value from continuum fitting with an additional QPO feature $p_\mathrm{AIC}$, quality factor $Q$, and fractional RMS amplitude.}\label{tab:qpo}

\end{table*}

\begin{figure}
    \centering
    \begin{subfigure}{\linewidth}
        \centering
        \includegraphics[width=\linewidth]{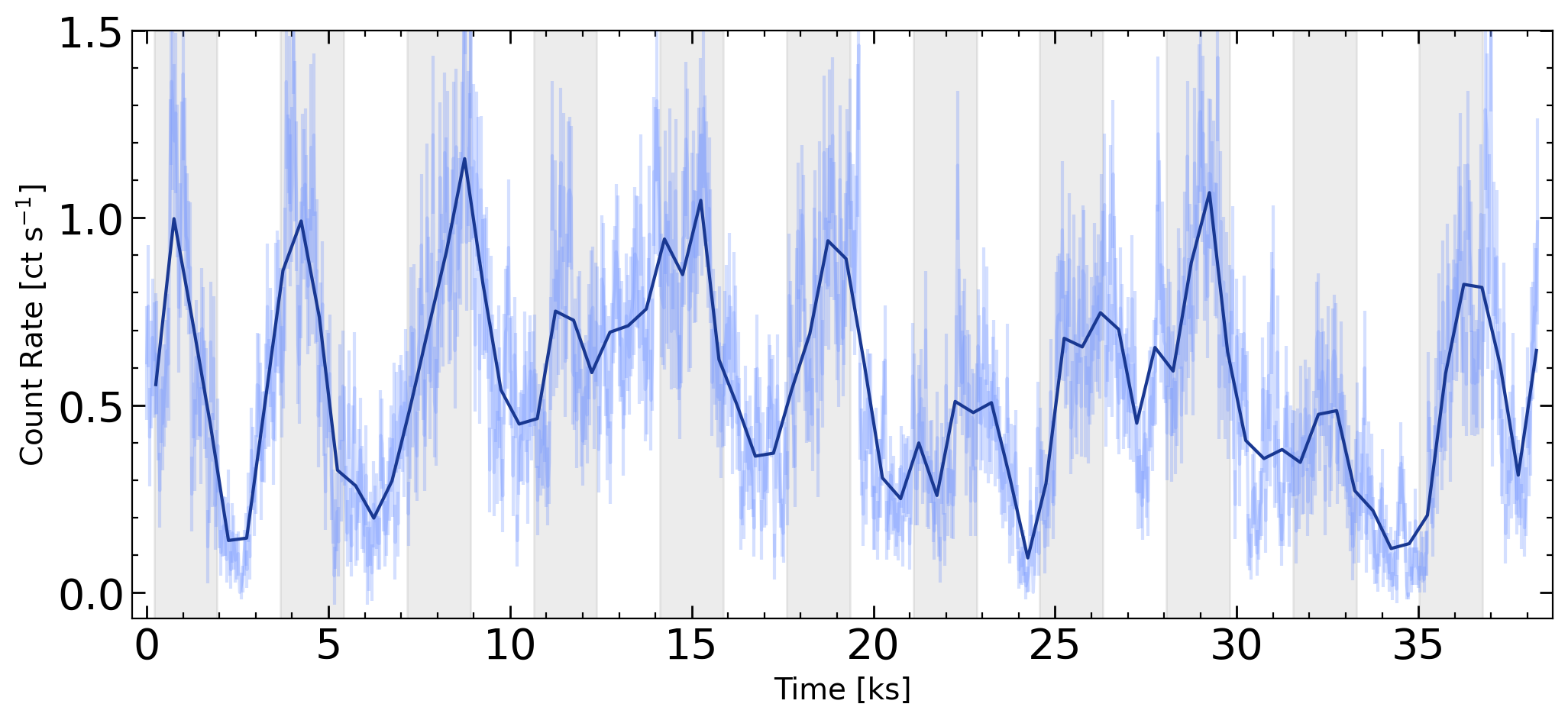}
        \label{fig:qpo_lc}
    \end{subfigure}
    \vspace{0.3cm}
    \begin{subfigure}{\linewidth}
        \centering
        \includegraphics[width=\linewidth]{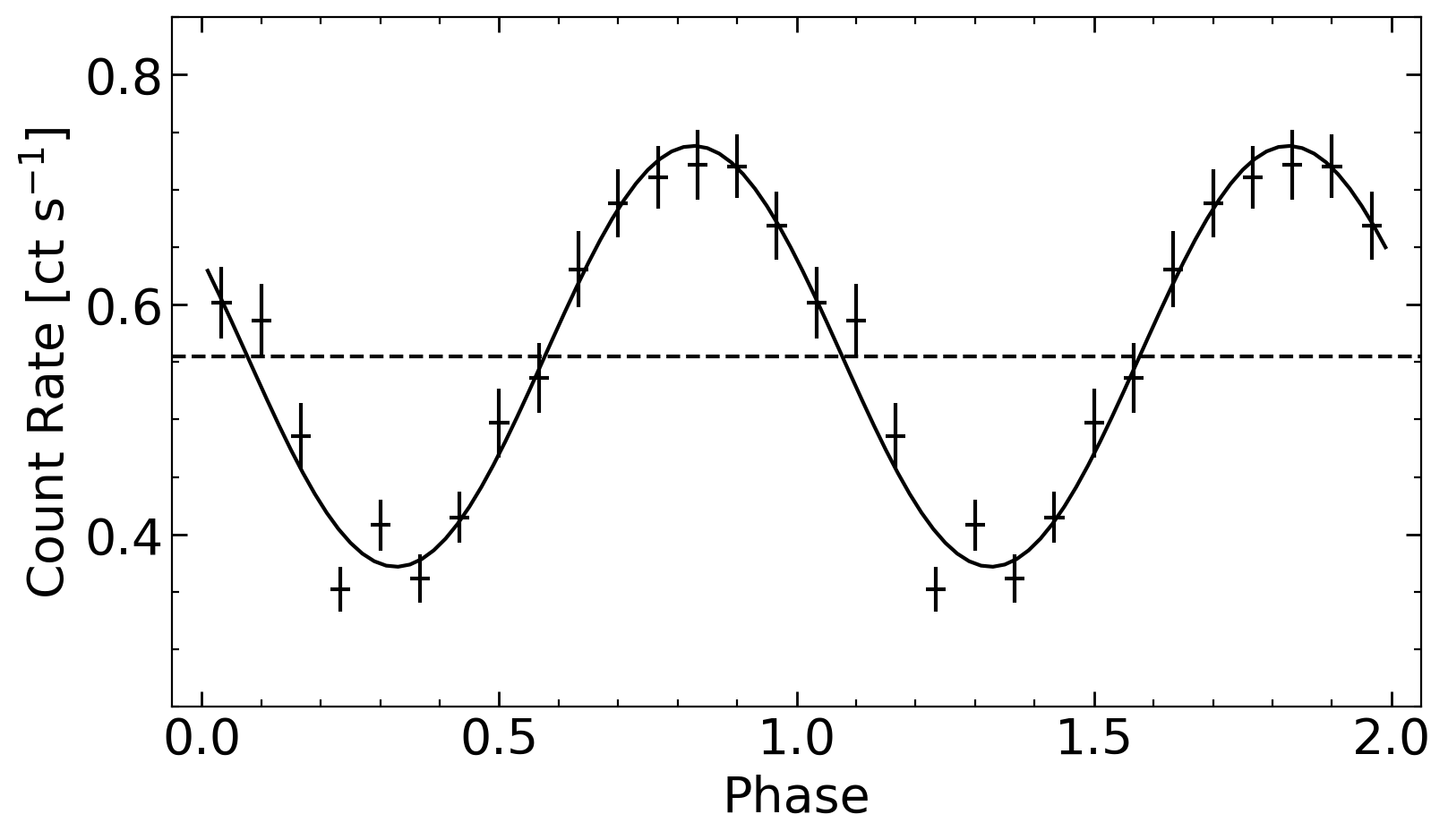}
        \label{fig:phase_folded_lc}
    \end{subfigure}
    \caption{\textit{Upper:} 0.3--7 keV light curve for XMM2. Gray shaded regions indicate alternating half-periods ($P= 3448$~s).
    \textit{Lower:} Phase folded light curve and best-fit sinusoid. 
    The light curve is binned at 15 bins per cycle, and two cycles are shown for visual clarity.}
    \label{fig:qpo_lc}
\end{figure}

\section{Discussion}\label{sec:discussion}

\subsection{Mass constraints}
Black hole mass estimation for tidal disruption events poses a challenge because TDEs lack a virialized broad line region (BLR), so standard AGN techniques based on BLR line widths are not applicable.
Since there is an upper limit to the mass at which a black hole can be observed to tidally disrupt a star, TDEs naturally probe the lower mass end of the SMBH mass function \citep{hills75, rees88, vanvelzen18}.
We often rely on host galaxy scaling relations, which can carry large uncertainty, particularly at lower mass \citep[e.g.,][]{gultekin09, kormendy13, reines15, yao23}.
More recently, attempts to extract mass from UV/optical light curve properties have shown promise \citep{mockler19, mummery23}.
Here we consider three mass estimation methods adapted from the XRB and AGN communities that leverage X-ray spectral and timing properties: (1) spectral continuum fitting with the relativistic accretion disk model \kerrbb, (2) the excess variance $\rms$, and (3) the QPO period.
We discuss the applicability and limitations of these X-ray methods for TDEs.
We find that the resulting mass estimates are consistent within $2\sigma$.
In Section \ref{sec:disc_lit_comp}, we compare our results with mass estimates from optical/UV modeling and host galaxy scaling relations.

\subsubsection{\kerrbb}\label{sec:disc_kerrbb}
The \kerrbb\, model considers blackbody continuum emission from a general relativistic accretion disk \citep{li05}. 
\kerrbb \, was developed for use with X-ray binaries, which often have independent black hole mass estimates.
Since accretion disk models are understood to constrain two parameters well, these independent mass estimates allow disk models to constrain spin and accretion rate for a fixed mass.
In this work, we make the assumption of a maximally spinning black hole to in turn estimate a mass and characterize an evolving accretion rate.

Our analysis is limited by some parameter degeneracies. 
We identify a correlation between the mass and the \ztbabs\, nH parameter, which is reduced when nH is tied across observations. 
MCMC analysis also reveals a degeneracy between \mbh\, and \mdot; however, the mass remains well constrained within the allowed range of \mdot. 
Because the X-ray emission does not necessarily trace the bolometric luminosity, we do not place strong weight on the inferred \mdot\, values.

Opacity effects in the disk atmosphere shift the emergent spectrum to higher energies. 
The color correction $\fcol=T_\mathrm{col}/T_\mathrm{eff}$ quantifies this hardening relative to blackbody emission
\citep{shimura95}.
The color correction factor is challenging to constrain because it encapsulates effects from multiple physical processes (electron scattering, emission/absorption opacity), is degenerate with other spectral parameters, and may vary over time \citep[e.g.,][]{nowak08, reynolds13, salvesen21}. 
We choose $\fcol=2.4$ based on theoretical  disc models for a $10^6~\msun$ black hole with a high accretion rate \citep{ross92, done12}.
There is uncertainty surrounding the color correction factor, with a possible range of $\approx1.4-2.5$ \citep[e.g.,][]{li05}. 
We explore this range and find that lowering \fcol\, only decreases the mass estimate, by at most a factor of three for the lowest tested value of $\fcol=1.4$.

We also freeze the disk inclination $(i=45\degree)$ and black hole spin $(a=0.999)$.
The disk inclination is positively degenerate with mass, resulting in mass estimates between $9.5\times10^4$ and $1.2\times10^6$ over the allowed inclination range of $0\degree-85\degree$. 
The spin correlates positively with inferred mass and negatively with accretion rate. 

The posterior black hole mass distribution from \texttt{xspec\_emcee} is shown in Figure~\ref{fig:mass} (dark red dashed line). The resulting estimate is the lowest but consistent with the other two X-ray based methods.

\subsubsection{Excess variance}\label{sec:disc_rms}
It has long been recognized that many of the physical timescales determining AGN variability (i.e. viscous, dynamical, light travel timescales) scale linearly with mass \citep[e.g.,][]{shakura73}. Therefore, less massive black holes will show more variability over a fixed duration \citep{lu01, bian03}. This result was quantified in \citet{ponti12}, who studied short-term X-ray variability in 161 AGN and found a strong negative correlation between the normalized excess variance $\rms$ in coronal emission (2--10~keV) and black hole mass.

Unlike AGN where the corona dominates, most TDEs show soft, disk-dominated X-ray spectra.
AT2019teq's hard state provides a unique opportunity to compare with coronal variability in AGN, assuming the same underlying physics.
We estimate the excess variance from the final three \textit{XMM-Newton} observations, which probe the TDE's corona-dominated state.
Using this variance, the black hole mass suggested by the \citet{ponti12} relation is $6.4 \pm 2.4 \times 10^5~\msun$.

The relation is less well constrained at low mass: only five AGN in the sample with reverberation mapped-masses have $\mbh < 10^7~\msun$.
Most of these lie below the best-fit line (i.e., showing lower $\rms$ for their mass), which suggests the trend may flatten at low mass.
We note that AT2019teq exhibits higher-amplitude variability than the reverberation-mapped AGN at comparable masses.

TDE light curves are nonstationary, and the observed excess variance decreases between XMM3,4 ($0.31-0.36$) and XMM5 ($0.09-0.13$).
Given the limited number of observations, it is unclear whether this reflects a long-term decrease in variability.
The nearly identical spectra across the three observations suggest they sample the same state.
That said, the lowest measured excess variance $\rms=0.09$ in XMM5 provides a strong upper limit on the black hole mass of $1.37 \times 10^6~\msun$. 

\subsubsection{QPO period}
Quasi-periodic oscillations can provide independent constraints on the black hole mass under the  assumption that QPOs are associated with orbital motion near the innermost stable circular orbit (ISCO).
Black hole mass can be computed directly by associating the QPO period with a Keplerian orbital period. Following \citet{gezari21}, the corresponding orbital radius is
\begin{equation}
    R_\mathrm{in} = \left( \frac{GM_\mathrm{BH}P^2 }{4\pi^2}\right)^{1/3} \sim 0.1 \left( \frac{P}{[100~\text{sec}]}\right)^{2/3}M_6 R_S
\end{equation}
The black hole mass is then
   \begin{equation}
     M  \sim 10 \frac{R_{\rm in}}{R_S} \left( \frac{P}{[100~\text{sec}]}\right)^{-2/3}\times10^6.
   \end{equation} 
We take $R_\mathrm{in} = R_{\rm ISCO}$, which depends on the spin $a$.  
For a maximally prograde spinning  black hole ($a=0.998$),  $R_{\rm ISCO} \approx 0.5 R_S$, and so $M  = 4.72 \pm 0.80 \times 10^5~\msun$. 
As spin decreases, the ISCO moves outward, and the corresponding mass increases.
Because the spin is not well constrained, we adopt the mass estimate based on maximal spin, which is broadly consistent with high spins observed in AGN \citep{brenneman13, vasudevan16, reynolds19} and allows for direct comparison with our \kerrbb\, estimate. Marginalizing over the full range of prograde spins gives a median black hole mass of $1.99\pm0.59\times10^6$~\msun.

QPOs may originate at larger radii than the ISCO \citep[e.g., Type-C QPOs in XRBs,][]{ingram09}.
Maintaining a fixed orbital frequency at a larger radius requires a smaller black hole mass.
Therefore, associating the QPO with the ISCO provides an upper limit on the black hole mass.

\subsubsection{Comparison to literature}\label{sec:disc_lit_comp}
Here we compare the masses inferred from our three X-ray techniques to those more commonly used in the TDE literature.

MOSFiT TDE modeling attributes UV/optical emission to X-ray reprocessing and assumes rapid circularization  \citep{mockler19}. 
MOSFiT assumes that optical luminosity follows the mass fallback and accretion rate, but this assumption may not hold for TDEs with late-time UV/optical plateaus. 
To mitigate this, TDE light curves may be truncated to early epochs for fitting.
As visible in Figure \ref{fig:lcs_all}, the ZTF-$g$ light curve of AT2019teq plateaus within $<300$~days of optical peak. 
\citet{hammerstein23} fitted  light curves to 300 days past optical maximum (MJD 59094, see Figure \ref{fig:lcs_all}) and obtained a black hole mass of $\log(\mbh/\msun)=6.05\pm0.4$. 
In contrast, \citet{alexander25} fit to only 56 days past maximum (MJD 58850) and estimated a lower mass of $\log(\mbh/\msun)=5.8\pm0.3$. 
We continue with this updated value from exclusively pre-plateau data. 
Our estimates are broadly consistent with MOSFiT modeling, though we note that the MOSFiT mass is highly sensitive to the choice of light curve truncation.

TDEMass uses shocks caused by small-angle apsidal precession to explain UV/optical emission rather than X-ray reprocessing, and assumes slow circularization \citep{ryu20}.
\citet{hammerstein23} reports an estimated mass of $\log(\mbh/\msun)=6.30\pm0.05$ using TDEMass.

Finally, we consider a host galaxy scaling relation. 
We lack a stellar velocity dispersion measurement for the host galaxy, and so we use an empirical linear relation between $\mbh$ and $M_\mathrm{gal}$ for TDEs developed by \citet{yao23}.
The resulting mass is $\log(\mbh/\msun)=6.32\pm0.49$, the largest of all estimates quoted here, but with significant uncertainty.

\begin{figure*}
    \centering
    \includegraphics[width=0.8\linewidth]{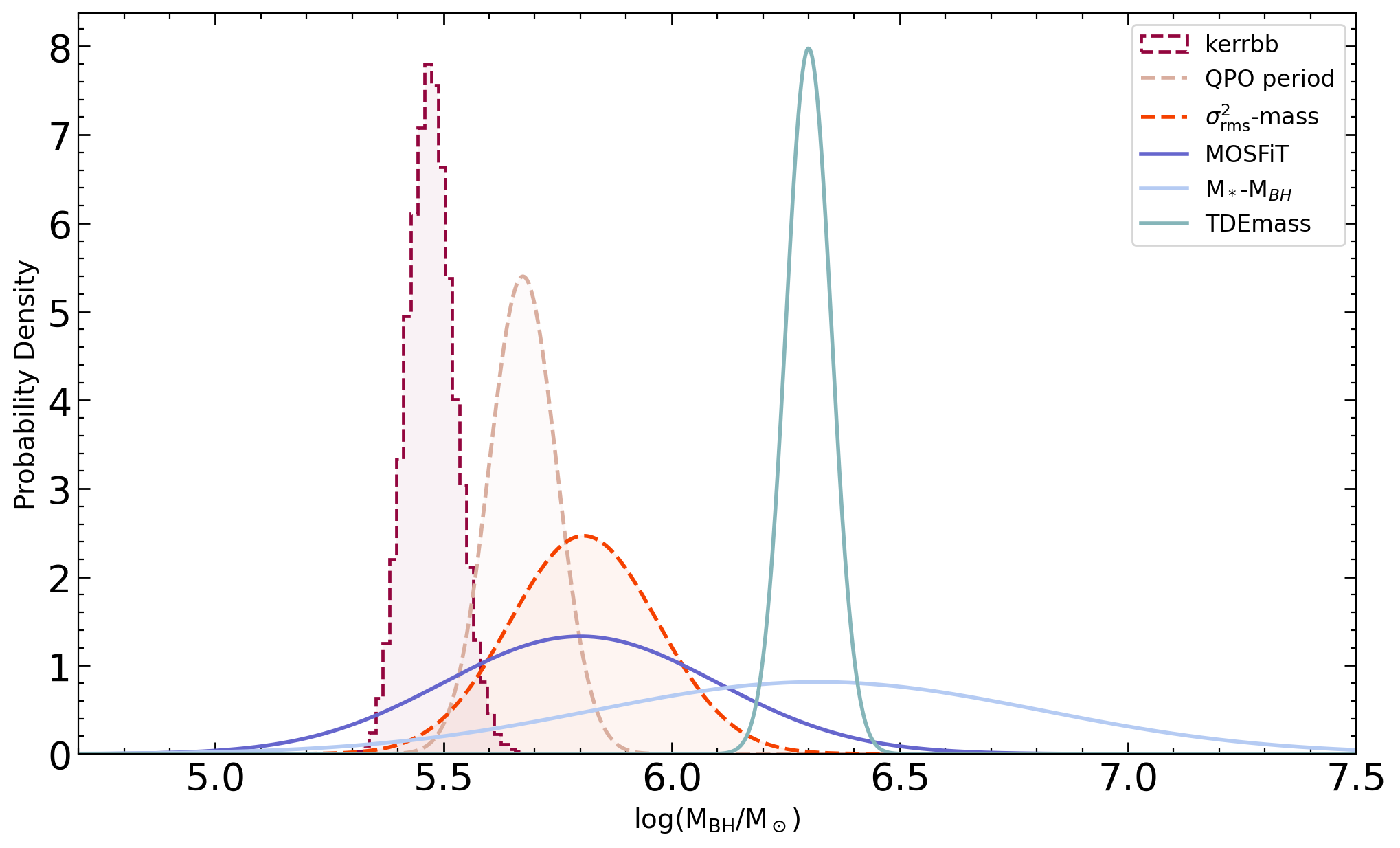}
    \caption{Constraints on black hole mass. 
    \textit{This work, dashed lines:} Posterior distribution of \kerrbb \, mass from MCMC analysis (dark red), Gaussian distribution from QPO period (beige), and 2--10~keV $\rms$-mass \citep[orange,][]{ponti12}.
    \textit{Literature, solid lines:} Gaussian distribution from MOSFiT \citep[purple][]{guillochon18, mockler19, alexander25}, TDEMass \citep[teal,][]{ryu20, hammerstein23}, and $M_{\mathrm{gal}}$--$M_{\mathrm{BH}}$ \citep[light blue,][]{yao23, guolo24}. 
    In general, the estimates from this work using X-ray properties suggest a lower black hole mass than those from optical/UV light curve modeling and host galaxy scaling relations.}
    \label{fig:mass}
\end{figure*}

Figure \ref{fig:mass} displays the constraints on black hole mass from each method described above. For consistency across methods we plot results for the assumption of maximal prograde spin. In general, the estimates from X-ray properties in this work predict a lower mass than the literature estimates based on optical/UV properties: the mean mass from this work is $\log(\mbh/\msun)=5.67\pm0.06$, while the mean mass from the three literature estimates is $\log(\mbh/\msun)=6.14\pm0.19$.

The mean mass estimate from this work is consistent with  the mean literature estimate to within $2.3\sigma$.
Our estimate is consistent with the MOSFiT value to within $<1\sigma$, but inconsistent with the TDEmass value.
The X-ray based estimates from this work are consistent with each other to $2.2\sigma$.
Our results suggest that AT2019teq may be less massive than previous literature estimates.

\subsection{State transitions}\label{sec:disc_state_transitions}
AT2019teq exhibits hard X-ray emission for at least 1100 days, indicating the longest-lived corona observed in a TDE.
Spectral hardening  in TDEs has been interpreted as the emergence of a magnetically supported corona on $\sim$100–200 day timescales \citep{yao22, guolo24}.
This corona formation may be driven by an amplified magnetic field due to differential rotation of the accretion disk and the magnetorotational instability \citep{balbus91, miller00}.
Late-time spectral hardening has been seen in AT2018fyk \citep{wevers21}, AT2020ocn \citep{cao24}, AT2021ehb \citep{yao22}, although in each case the X-ray emission appeared to soften or faded within a few hundred days of optical peak.
AT2019teq's sustained high X-ray luminosity and persistent hard state provide a unique opportunity to investigate long-term accretion state evolution in massive black holes.

The potential scale-invariance of accretion physics has received considerable attention, e.g. in comparative studies spanning 6-8 orders of magnitude in black hole mass \citep[e.g.,][]{mchardy06}.
AT2019teq exhibits several similarities to XRB and AGN accretion states. 
Its late-time radio brightening around the onset of the hard state after a non-detection in early times aligns with both  XRBs \citep{gu09} and AGN \citep{kording06, svoboda17} for being radio-louder in hard states.

\begin{figure}[ht]
    \centering
    \includegraphics[width=0.95\linewidth]{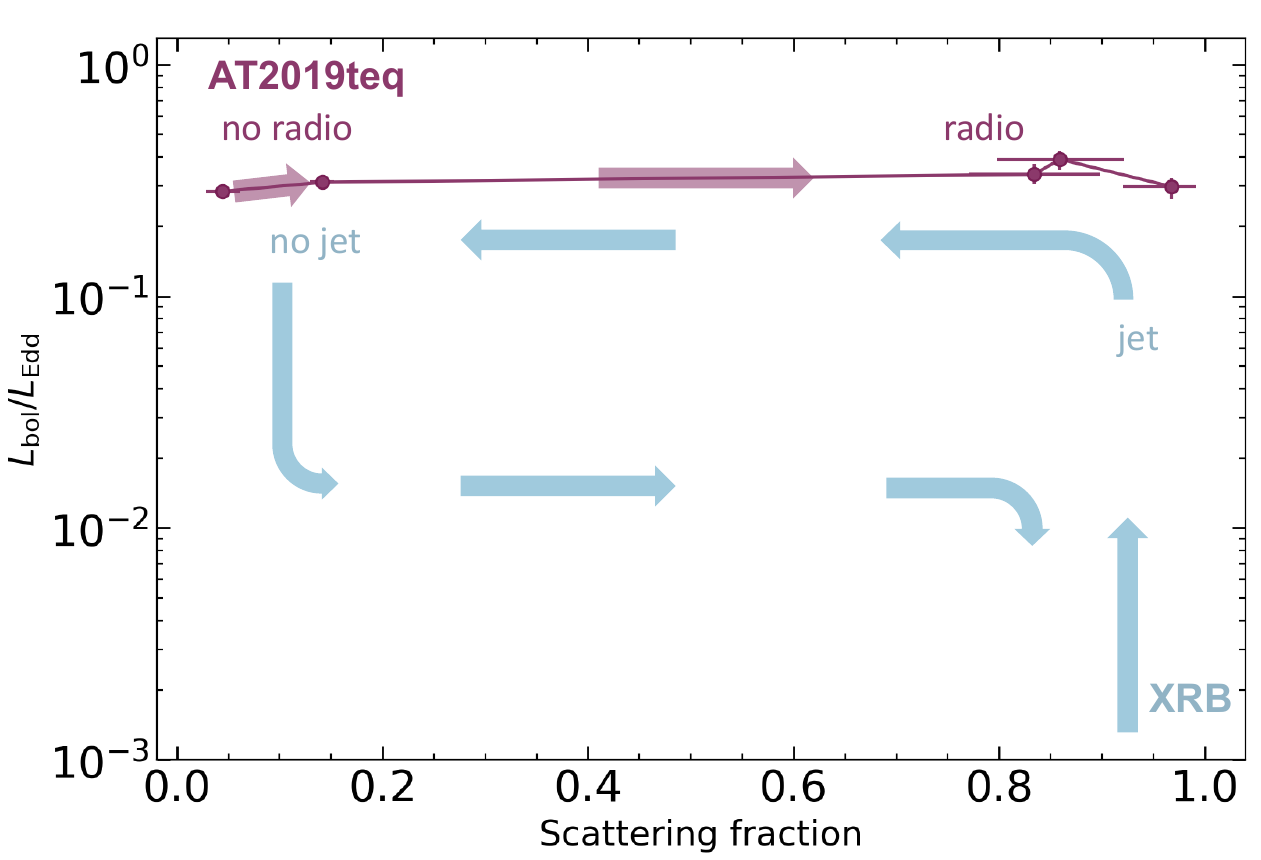}
    \caption{$L_\mathrm{bol}/L_{\mathrm{Edd}}$ vs scattering fraction. 
    A schematic X-ray binary trajectory is overplotted where $L_\mathrm{bol}\approx L_\mathrm{X}$. Arrows indicate the direction of time.}
    \label{fig:hid}
\end{figure}

State transitions are common in X-ray binary outbursts  \citep{remillard06, wang22}.
The schematic hardness--intensity diagram in Figure \ref{fig:hid} illustrates AT2019teq's trajectory against the canonical q-shaped evolution of X-ray luminosity as a function of coronal strength.
An XRB system evolves from a bright hard state, typically associated with coronal emission and a mildly relativistic jet, to a bright soft, disk-dominated state, then hardens after dimming.
While X-ray luminosity dominates the bolometric output of XRBs, the UV/optical component dominates TDE emission at early times. 
We thus estimate AT2019teq's bolometric luminosity by combining the 0.3--10~keV X-ray luminosity with a blackbody-derived bolometric luminosity from \citet{hammerstein23}. The blackbody component is reconstructed from their best-fit parameters for a gaussian-rise, power-law-decay model applied to the optical (ZTF $g$ and $r$) and UV  (Swift UVOT) data.

AT2019teq's bolometric luminosity remains roughly constant as the UV/optical flare fades and X-ray emission strengthens.
The Eddington ratio $L_\mathrm{bol}/L_{\mathrm{Edd}}\approx0.3$ is consistent with values observed in other TDEs \citep[0.01-1,][]{mummery23}.
TDEs have been interpreted as XRB soft-state analogues based on both this range of Eddington ratios and their soft spectra \citep{mummery23}.
However, AT2019teq's spectral hardening complicates such an interpretation.
In XRBs, Eddington ratios $\gtrsim0.01$ typically correspond to high states where sources transition from hard to soft, in contrast to AT2019teq's evolution.
A clear interpretation is additionally complicated by the fact that early optical TDE emission likely originates from reprocessing or stream-stream collisions, rather than the disk itself.
These properties suggest that while AT2019teq shares similarities with XRB accretion states, its spectral evolution shows the mapping between TDE and XRB state transitions may not be straightforward.

While AT2019teq's accretion states present compelling analogues to XRBs, its soft-to-hard transition occurs on a timescale far shorter than expected from linear mass scaling.
The transition to a corona-dominated state occurred within three years; assuming a roughly linear scaling with black hole mass, this corresponds to tens of minutes to a few hours on $10~\msun$ XRB timescales.
If state transitions in XRBs and TDEs are driven by the same underlying mechanism, such a rapid transition would imply unusually efficient coronal formation, perhaps facilitated by a faster buildup of magnetic flux.
Alternatively, the source may not represent a fully canonical hard state; its relatively soft photon index $\Gamma=2.84$ is consistent with this interpretation.
Should AT2019teq remain bright, continued monitoring could uniquely constrain the long-term evolution of a corona in a TDE.

State transitions are rarely observed in AGN, as they are expected to occur over timescales far exceeding  observational baselines.
An exception to this is Changing-Look AGN (CLAGN), which exhibit the appearance or disappearance of broad lines driven by changes in obscuration or accretion state \citep[e.g.,][]{shappee14, ricci23}.
Notably, the CLAGN 1ES 1927+654 provided the first observational evidence of corona formation in an AGN \citep[][]{ricci20, masterson22}, with the corona forming on a timescale ($\sim$years) comparable to AT2019teq.

\section{Conclusions}\label{sec:conclusions}
In this work, we have analyzed the spectral and timing properties of the tidal disruption event AT2019teq. Our main results are as follows.
\begin{enumerate}[label=\roman*.]
    \item AT2019teq exhibits a state transition from a soft disk-dominated state to a hard corona-dominated state (Figure \ref{fig:spectra}).
    The hard state has persisted for at least 1100 days, indicating the longest-lived corona known in a TDE.
    The state change enables disk- and corona-based estimates of the black hole mass.
    \item AT2019teq begins in a disk-dominated state. If we associate this with a disk that extends to the ISCO for a maximally prograde spinning black hole, we infer a mass of $3.0^{+0.4}_{-0.3}\times10^5$~\msun.
    
    \item  After $\gtrsim100$ days AT2019teq transitions to harder, corona-dominated state. Under the assumption of similar physics between the corona in TDEs and AGN, we  use the variability timescale as a probe of black hole mass and infer a mass of $6.4 \pm 2.4 \times 10^5\,\msun$.

    \item We detect a tentative sub-mHz QPO in one epoch with a period of 3448~s/$2.9\times 10^{-4}$ Hz (Figures \ref{fig:qpo_psd} and \ref{fig:qpo_lc}).
    By associating the QPO with emission from the ISCO of a maximally rotating black hole, we infer a black hole mass of $M  = 4.72 \pm 0.80 \times 10^5~\msun$.
    
    \item
    We find that all three X-ray-based mass estimates are consistent with each other, and provide slightly lower masses (mean $\log(\mbh/\msun)=5.67\pm0.09$) than those in the literature from optical light curve modeling and galaxy scaling ($\log(\mbh/\msun)=6.14\pm0.19$).
\end{enumerate}

The physics of the X-ray emission in tidal disruption events is in some ways better understood than at longer wavelengths, and X-ray based measurements of the black hole mass offer a promising path forward. 
Accurate black hole masses are essential for testing scale-invariant accretion; our work reveals striking parallels between TDE and XRB accretion state properties, though AT2019teq's relatively rapid spectral evolution raises questions about distinct transition timescales in TDEs.
Thus, long-term monitoring and systematic timing studies of individual sources will be crucial to characterize state transitions and variability evolution in TDEs.
Upcoming surveys such as Rubin Observatory’s Legacy Survey of Space and Time \citep{ivezic19} and ULTRASAT \citep{shvartzvald24} will discover unprecedented numbers of TDEs, enabling both larger population studies and targeted follow-up.
Particularly promising targets for timing studies are TDEs around lower-mass black holes, where QPO periods fall on observable timescales, and hard-state TDEs. 
These would help constrain coherent variability behavior in TDEs and enable more direct comparisons with XRB and AGN variability properties to advance our understanding of accretion physics across black hole mass scales.

\section*{Acknowledgments}
VB was supported by the NSF GRFP under Grant \#2141064.

\bibliography{bibliography}
\bibliographystyle{aasjournal}

\appendix
\counterwithin{figure}{section}
\counterwithin{table}{section}

\section{High-energy background light curves}\label{app:bg}

\begin{figure}[h]
    \centering
    \includegraphics[width=\linewidth]{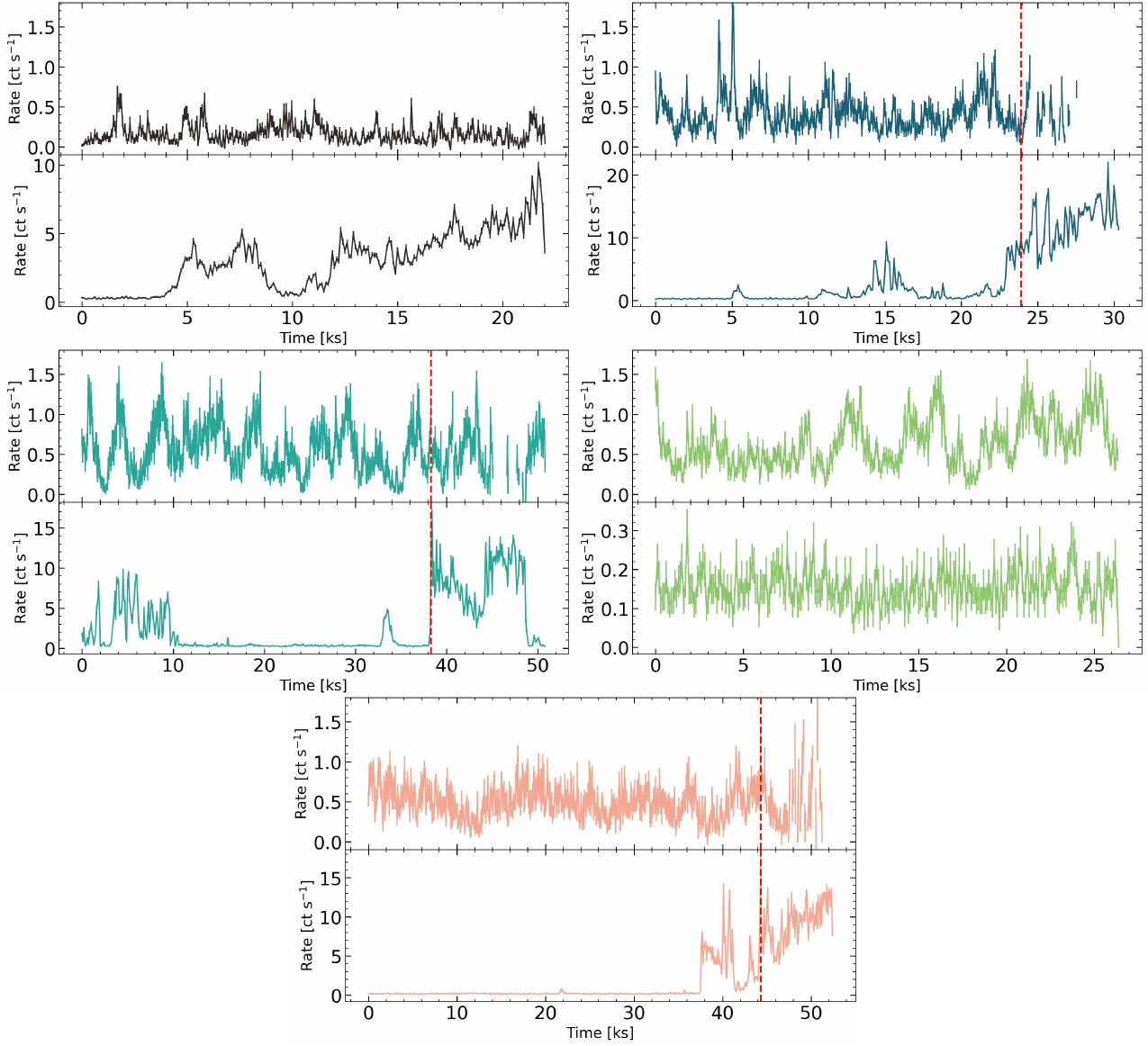}
    \caption{\textit{XMM-Newton} background-subtracted science light curves, and 10--12~keV particle-background light curves. The dashed red lines mark the times beyond which the high-energy background remains above 10~ct~s$^{-1}$; intervals beyond these times are discarded.}
    \label{fig:placeholder}
\end{figure}

\end{document}